\documentclass[twocolumn,showpacs,preprintnumbers,amsmath,amssymb,prl,floatfix]{revtex4}

\usepackage{psfig}
\usepackage{epsfig}
\expandafter\ifx\csname package@font\endcsname\relax\else
 \expandafter\expandafter
 \expandafter\usepackage
 \expandafter\expandafter
 \expandafter{\csname package@font\endcsname}%
\fi
\begin{document}


\title{Coexistence of glassy antiferromagnetism and giant magnetoresistance (GMR)
\\in Fe/Cr multilayer structures }

\author{N. Theodoropoulou}
\author{A. F. Hebard}%
 \email{afh@phys.ufl.edu}
\affiliation{%
Department of Physics, University of Florida, Gainesville, FL 32611-8440\\
}%
\author{M. Gabay}%
\affiliation{%
Laboratoire de
Physique des Solides, Bat 510, Universite Paris-Sud, 91405 ORSAY Cedex, France\\
}%
\author{A.K. Majumdar}%
\affiliation{%
Department of
Physics, Indian Institute of Technology, Kanpur-208016, India\\
}%
\author{C. Pace}%
\author{J. Lannon}%
\author{D. Temple}%
\affiliation{%
MCNC, Electronics Technologies
Division, Research Triangle Park, NC 27709\\
}%
%
\date{\today}

\begin{abstract}
Using temperature-dependent magnetoresistance and magnetization
measurements on Fe/Cr multilayers that exhibit pronounced giant magnetoresistance
(GMR), we have found evidence for the
presence of a glassy antiferromagnetic (GAF) phase.
This phase reflects the influence of interlayer exchange coupling (IEC) at low
temperature ($T < 140$K) and is characterized by a field-independent glassy
transition temperature, $T_g$, together with irreversible behavior having logarithmic time
dependence below a ``de Almeida and Thouless'' (AT) critical field line.
At room temperature, where the GMR effect is still robust,
IEC plays only a minor role, and it is the random potential variations acting on the
magnetic domains that are responsible for the antiparallel interlayer domain alignment.

\end{abstract}

\maketitle

Given the established presence of GMR-based devices in technology, especially
in the multi-billion dollar computer hard disk drive market, it may come as a surprise
that there is still an incomplete scientific understanding of the GMR effect\cite{Har99}. The
mechanism for GMR, first observed in single crystalline (100) Fe/Cr multilayers grown
by molecular beam epitaxy\cite{Bai882472,Bin894828} and subsequently in
magnetron-sputtered polycrystalline
films\cite{Par902304}, relies on spin-dependent scattering\cite{Fer76849}
and the associated dependence of resistance
on the relative orientations of the magnetizations in neighboring layers. It is important to
recognize that interlayer exchange coupling (IEC) is not necessarily required for a GMR
effect\cite{Har99}.
In a particularly simple manifestation, two neighboring films, separated by a
non-magnetic spacer layer, could have different coercive fields, thus giving rise to
antiparallel alignment and a GMR effect, as the external field is cycled\cite{Bar908110}.
Randomness\cite{Zha99468,Mil004224}
and competing interactions such as biquadratic coupling\cite{Slo9413,Fil951847}
can also play a significant role.
In this paper we identify a glassy antiferromagnetic (GAF) phase which by marking the influence of IEC
at low temperatures implies that at higher temperatures random potential variations rather than IEC 
are responsible for antiparallel alignment.

Our Fe/Cr multilayer samples have been prepared on silicon substrates by ion
beam sputter deposition of separate Fe and Cr targets. Extensive characterization of the
deposited multilayers showed distinct compositional and structural modulations with
well-defined interfaces and a surface roughness on the order of 5\AA. Ten and thirty-layer
stacks with the repeat sequence [Fe(20\AA)/Cr($d_{Cr}$)] are typically deposited and passivated
with a 50\AA-thick Cr layer. The Cr spacer thickness $d_{Cr}$ is varied over the range 8--12\AA.
The inset of Fig.~1 shows typical GMR traces at 300K and 10K for the magnetic field
parallel to the planes of a [Fe(20\AA )/Cr(12\AA ) ]$\times$30 sample.

In Fig.~1 we show a selected subset of temperature-dependent field-cooled (FC,
open symbols) and zero-field-cooled (ZFC, closed symbols) magnetization data for a
thirty layer sample with $d_{Cr} = 12 \AA$ and a GMR ratio ($(R(0)-R(H))/R(0)$, Fig.~1 inset)
of 20.6\% at 10K.
The data were taken using a SQUID magnetometer in fields (indicated on the plot) oriented
parallel to the layers.
At each field the corresponding FC and ZFC curves can be
characterized by three distinct temperatures: an irreversibility temperature $T_{irr} (H)$
denoting the bifurcation point below which there is hysteresis (upward arrows), a
temperature $T_m (H)$ (downward arrows) denoting the maximum in each of the ZFC
curves, and an inflection temperature $T_{infl}$ (vertical dashed line) which marks the
inflection point of each FC curve. Evidently $T_{infl}$ is quite robust and independent of
field, having a value $T_{infl}= 93.0 \pm 1.4K$ determined to relatively high precision from FC
measurements at 5 different fields spanning the range 50-400 Oe.

\begin{figure}[t]
\begin{center}
      \includegraphics[width=0.9\linewidth]{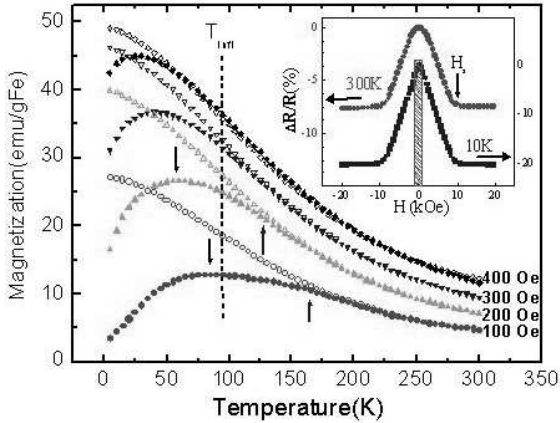}
\caption{\label{fig:fig1}
Magnetization of a multilayer sample ([Fe(20\AA)/Cr(12\AA)]$\times$30) normalized to
the weight of iron plotted as a function of temperature at the indicated fields. The data at
each field are taken in pairs: the open(solid) symbols referring to the field-cooled, FC,
(zero-field cooled, ZFC) procedure.
The vertical arrows and dashed line are described in the text.
Inset, dependence of the giant magnetoresistance (GMR) ratio on applied field for the same
film at 300K (left axis) and at 10K (right axis).
}
\end{center}
\end{figure}

Compelling evidence for an interlayer rather than intralayer effect is found in the
resistance measurements of Fig.~2 on the same sample.  For each datum on this graph,
the sample was zero-field cooled to the target temperature, the resistance $R(0)$ measured,
and then a field applied to measure the change in resistance $\delta R  =  R(0) - R(H)$.
The ratio $\mid \delta  R/R(0) \mid$ is plotted against temperature for the fields indicated
in the legend. The
striking aspect of these data is that although the peaks are not as pronounced as those in
the ZFC magnetizations of Fig.~1, their positions in an $H$-$T$ plot of Fig. 3 (open
triangles) show close similarity with respect to the positions of the ZFC peaks (solid
circles).

\begin{figure}[bh]
\begin{center}
      \includegraphics[width=0.9\linewidth]{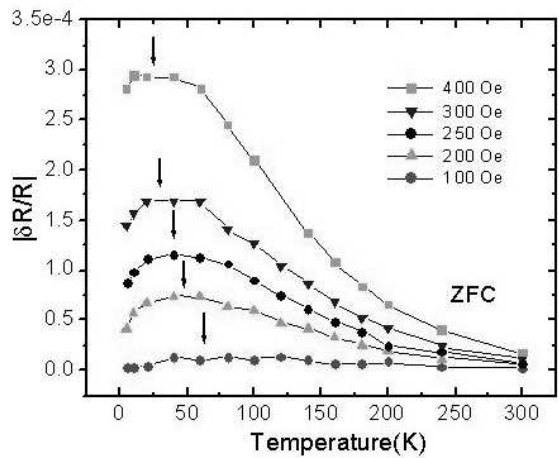}
\caption{\label{fig:fig2}
Temperature dependence of the relative changes in resistance at the fields
indicated in the legend for the same sample characterized in Fig.~1. For each data point,
the sample was zero-field cooled as described in the text.
The vertical arrows indicate the positions of the maxima for each field
and define a critical field dependence similar to that defined by the maxima of the ZFC
magnetizations in Fig.~1.
}
\end{center}
\end{figure}

The presence of a spin-glass-like phase is buttressed by our finding that $T_m(H)$
defines a critical field line (solid circles in Fig.~3) which delineates the onset of
strongly irreversible behavior and has the de Almeida and Thouless (AT)
form\cite{Alm78983,Bin86801}, $H/T \propto (T_g/T-1)^{3/2}$
(inset), where $T_g$ is the spin glass temperature. Although other criteria could have been
used\cite{Bin86801}, we note that our choice of $T_m(H)$ as the criterion determining
the AT line has
particular cogency because it obeys the scaling form of the AT prediction and
extrapolates at zero field to a field-independent glass temperature
$T_g = 1.51 \times T_{infl} =140$K,
where $T_{infl}$, an apparent fixed point, has been independently determined from the FC
data (dashed line of Fig.~1).

\begin{figure}[b]
\begin{center}
      \includegraphics[width=0.9\linewidth]{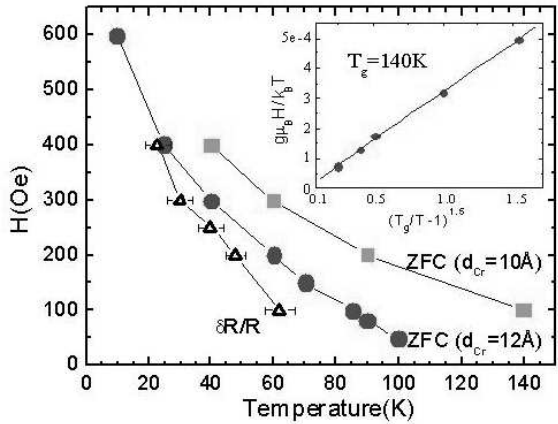}
\caption{\label{fig:fig3}
Critical field lines for the 30 layer [Fe(20\AA)/Cr(12\AA)]  (solid circles and open
triangles) sample shown in Fig.~1 and for a second 30 layer [Fe(20\AA)/Cr(10\AA)] (solid
squares) multilayer sample with smaller Cr spacer thickness. The solid symbols refer to
determinations using the experimental $T_m(H)$'s of ZFC magnetizations and the open
triangles are determined by similar peaks in the resistance measurements.
Inset, plot of the high temperature
points (solid circles) showing the de Almeida-Thouless (AT) scaling dependence for
spin glasses.
}
\end{center}
\end{figure}

An additional and essential ingredient for a glassy phase is the presence of
disorder measured by the variance, $\Delta J$, in the antiferromagnetic (AF) coupling strengths.
This variance arises because of the existence of domains and the concomitant
constraints imposed by intralayer dipolar interactions The exchange energy between two
Fe moments separated by a spacer layer is of the form $E= J_{AF} \textrm{cos} (\Psi)$,
where $\Psi$ denotes
their relative angle. The intralayer domain structure imposes well-defined orientations of
the spins and this constraint will not be consistent, in general, with $\Psi = \pi$ (i.e. with a
minimum value of $E$). Because of the long-range nature of dipolar interactions, lowering
the exchange energy requires the overturning of one or of several clusters of Fe
moments, which is energetically inhibited at low temperature. In this regime, $\Psi$ behaves
like a pseudo random variable. A realistic estimate for $\Delta J$ can be obtained by assuming a
flat distribution for the values of $\Psi$ on the [0, 2$\pi$] interval,
leading to $\Delta J = J_{AF}/ \sqrt{2}$.
At $T>T_g$, IEC is present but ineffective because the intralayer dipolar interactions dominate.

Many glassy systems, including the one discussed here, show AT like
boundaries without being Ising spin glasses to which the theory\cite{Bin86801,Cra82158}
strictly applies.
The GAF phase associated with our GMR multilayers is clearly not an Ising system and is
more reasonably described in terms of an anisotropic vector model in which the
elemental spins, belonging to magnetized domains, are coupled ferromagnetically in the
X-Y plane and antiferromagnetically in the perpendicular direction. For such vector
glass systems there is an additional degree of freedom in the order parameter and the
true phase boundary is delineated at higher temperatures and fields by the
Gabay-Toulouse (GT) boundary\cite{Gab81201}. A more comprehensive viewpoint that facilitates
understanding of our experiment can be gleaned from the schematic phase diagram,
shown in Fig.~4 for the $H$-$T$ plane at $J_{AF}/ \Delta J>1$.
(Note that the PM phase is not labeled as a
ferromagnetic (FM) phase, since in the presence of a field there is no
spontaneous symmetry breaking as the temperature is reduced
through the Curie temperature.)
In simplified terms the GT line (solid)
can be thought of as denoting the onset of a phase transition to glassy behavior and the
AT line (dotted) as the onset of pronounced irreversibility. (The experimental signature
of the GT line, which has not been measured here, is a divergence in the transverse ac
susceptibility.)
At $H=0$ both lines terminate at $T=T_g$.

\begin{figure}[b]
\begin{center}
      \includegraphics[width=0.9\linewidth]{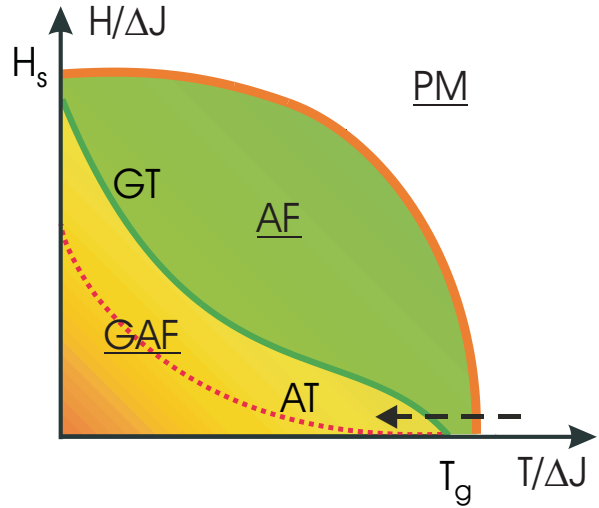}
\caption{\label{fig:fig4}
Schematic of phase diagram in the H-T plane showing the relationship
between the glassy antiferromagnetic (GAF), the antiferromagnetic (AF) and
the paramagnetic (PM) phases.
The axes are normalized as discussed in the text.
The Gabay-Toulouse (GT) and de Almeida and Thouless
(AT) line (dashed) are described in the text.
For our samples the disorder is
sufficiently large (i.e., $\Delta J \simeq J_{AF}$)
and the field sufficiently low to ensure that the
presence of an AF phase is obscured on the transition from the PM to GAF
phase (horizontal dashed arrow).
}
\end{center}
\end{figure}

The following three consequences, confirmed by experiment, are immediately
apparent: Firstly, since $T_g \propto J_{AF}$ and $\Delta J \simeq J_{AF}$,
it is clear that as $T_g$ increases, the
boundary of the GAF phase moves out to higher temperatures and fields.
Experimentally this is confirmed in Fig.~3 where the AT line for the sample with
$d_{Cr} = 10 \AA$ (solid squares) has higher critical fields and a correspondingly
higher $T_g$ than the sample with 12$\AA$ spacer.
A second consequence is that the disorder-induced close proximity of $T_g$
and $J_{AF}$ implies that at low $H$ the presence of an AF phase is obscured
on the transition (Fig.~4, horizontal dashed arrow) from the PM to GAF phase.
If this were not the case, then the field-cooled dc susceptibility
would have a maximum at the AF boundary and then saturate at a smaller value as
$T \rightarrow 0$.
Such maxima are not observed! A third consequence supporting the existence of a
GAF phase comes from the scaling of the field-cooled magnetization with $H$. Field-
cooled (FC) magnetizations including those shown in Fig.~1 reveal
that $M/H \sim H^{-u}$ as $T \rightarrow 0$.
Here we find $u$=0.58(2) for 5K magnetization data taken at 7 different fields
ranging from 100 to 800Oe, thus confirming behavior characteristic of spin glass
systems below the lower critical dimension \cite{Bin86801}.
Finally, in addition to hysteresis, we also
observe slow relaxations in the magnetization and resistance that are logarithmic in time
and which, but for lack of space, can be explained by invoking constraints on the
dynamics imposed by a hierarchy of domain sizes \cite{Gab866281,Pro996956}.

To fully appreciate the role of randomness in multilayers, it is important to
recognize the difference between GMR multilayers, in which there is a strong
interaction between closely coupled interfaces, and bilayer or trilayer configurations in
which such interactions can be ignored since there are at most only two interfaces. Thus
for example, in studies of exchange bias in single ferromagnetic/antiferromagnetic
(Co/CoO) bilayers\cite{Mil004224}, the onset of exchange bias,
which is induced by random interactions\cite{Zha99468},
is observed to occur at a single temperature, the Neel temperature.
By contrast, in our
case there are two temperature ranges: $T < T_g$ = 140K for glassiness and $T > \sim$250K
where there is a loss of AF order in Cr and disorder is still important. Accordingly, the
picture described for FM/AF bilayers\cite{Mil004224} is different for closely coupled
multilayers where
interactions between multiple ferromagnetic (FM) layers and interactions between
interfaces should be taken into account. Similar considerations also apply to the
magneto-optic Kerr effect (MOKE) and scanning electron microscopy with polarization
analysis (SEMPA) studies\cite{Pie99290} on Fe/Cr/Fe trilayers and magnetization
and ferromagnetic
resonance studies of CoFe/Mn/CoFe trilayers\cite{Fil951847},
all of which specialize to a specific type
of spacer layer and do not include the multilayer interactions responsible for our GAF
behavior. Our results are thus complementary yet distinct from the results of
bilayer/trilayer experiments.

A consideration of the relevant energy scales and the mutual interactions of the
magnetized domains in the Fe layers solidifies this emerging picture of spin-glass-like
behavior in GMR multilayers. If adjacent Fe layers of thickness $t$ and saturation
magnetization $M_s$ are coupled through an antiferromagnetic exchange $J$ per unit area,
then saturation at a field $H = H_s$ occurs when $J = HM_st/4$, a relation found by equating
the field energy per unit area, $HM_st$, to the energy difference, 4$J$, between the aligned
and antialigned magnetic configurations . We note that a glass temperature near 140K
corresponds to an antiferromagnetic coupling energy $\simeq 10$ meV, in good agreement
with theoretical calculations\cite{Fis9913849,Maj02054408} for Fe/Cr layers.
In the first calculation by Fishman and
Shi\cite{Fis9913849} the Fe layers are exchange coupled below the Neel temperature
$T_n$ of the Cr spacer
and a very strong AF coupling between the Fe and Cr moments at the interface is
assumed. For our GAF phase $T_n$ is in reality $T_g$. In the second calculation by Majumdar
\textit{et al}.\cite{Maj02054408} magnetoresistance data is well described by
a theoretical expression in which
RKKY interactions give a best fit AF coupling strength of $(70\pm20)$K.

For $T>T_g$, the Fe layers are no longer AF coupled and the expression
$J=HM_st/4$ to calculate the IEC is no longer relevant.
In its place we use the expression\cite{Koo607,Gab86655}
$H_s=4 \pi M_s$, to calculate the maximum saturation field necessary to align dipolar-coupled
domains within each layer. This expression is valid for both perpendicular and parallel
fields\cite{Gab86655}. The saturation fields of 10-20kOe in our samples (Fig.~1 inset) and similar
samples reported by others\cite{Bai882472,Par902304} are the right order of magnitude
for Fe which with a
saturation magnetization $M_s=1700$Oe/cm$^3$ implies a maximum saturation field
$H_s = 4 \pi M_s = 21$kOe. For our three different samples with
$d_{Cr}$ = 8, 10 and 12$\AA$ we find a linear
dependence of $H_s$ on $d_{Cr}$ which extrapolates to the origin ($d_{Cr}=0$)
to a value within 5\% of $H_s$=21kOe, thus validating our use of this analysis.

To associate field scales with energy (or equivalently, temperature), we use the
conversion ratio, 2.2$\mu_BB/k_BT$=1.5T/K, where the magnetic moment of Fe is 2.2 Bohr
magnetons. Accordingly, the dipolar interaction strengths measured by $H_s$, which are
balanced by domain wall energies, are on the order of a few Kelvin and hence not strong
enough at $T>T_g$ to determine domain orientation. Rather, domain orientation at $T>T_g$
is determined by the much stronger potential variations associated with crystalline
anisotropies and the presence of impurities and defects. The presence of a GAF phase
implies that IEC is effective in creating an anti-alignment effect beneficial to a large
GMR effect only at low temperatures ($T<T_g$) and low fields ($H<H_{AT}$). The shaded
region in the inset of Fig.~1 illustrates just how narrow this region is.

In summary, we show that a heretofore-unrecognized glassy antiferromagnetic
(GAF) state coexists with GMR in polycrystalline Fe/Cr multilayer stacks. The very
presence of this glassy phase sets an energy scale ($T_g$=140K) for antiferromagnetic
interlayer exchange coupling (IEC) that is well below room temperature. We therefore
conclude that, for temperatures greater than $T_g$, IEC plays only a minor role in forcing
the antiparallel interlayer domain orientations that give rise to the ($H=0$) high resistance
state of multilayer Fe/Cr GMR samples. Rather, random potential variations, which
constrain domain orientation, must be taken into account to understand GMR in
multilayer GMR devices. The origin of the dependence of $H_s$ on spacer thickness in
multilayers as observed here and by others\cite{Bai882472,Par902304} as well as the
origin of the AF couplings for
$T<T_g$ are totally open questions. This contrasts with the bilayer and trilayer
cases\cite{Zha99468,Mil004224,Pie99290} for
which the AF couplings have a clear source.

We thank S. B. Arnason, S. Hershfield, P. Kumar and C. Yu for valuable discussions and
suggestions.
This work was supported by AFOSR, DARPA and NSF.
\bibliography{final}
\end{document}